\documentclass[prd,notitlepage,longbibliography,nofootinbib,superscriptaddress,onecolumn,preprintnumbers]{revtex4-2}
\usepackage[utf8]{inputenc}

\usepackage{bm}
\usepackage{comment} 
\usepackage[colorlinks=true,urlcolor=blue,anchorcolor=black,citecolor=blue,linkcolor=red,filecolor=black,menucolor=black,pagecolor=black,linktocpage=true,pdfproducer=medialab,pdfa=true]{hyperref}
\usepackage{graphicx}
\usepackage{amsmath,latexsym,amssymb,mathrsfs,ascmac,mathtools}
\usepackage{multirow}
\usepackage{braket}
\begin{document}

\title{Wheeler–DeWitt Equation for Black Hole Interiors in Asymptotically Safe Gravity}

\author{Takamasa Kanai}
\email{kanai@kochi-ct.ac.jp}

\affiliation{Department of Social Design Engineering,
National Institute of Technology (KOSEN), Kochi College,
200-1 Monobe Otsu, Nankoku, Kochi, 783-8508, Japan}

\begin{abstract}
In this work, we analyze the Wheeler–DeWitt equation with scale-dependent gravitational couplings within the framework of asymptotically safe gravity. In the Hamiltonian formulation based on a renormalization-group improved Einstein–Hilbert action, the consistency of the theory and the Poisson algebra of constraints have been clarified. Within this framework, we show that, despite the explicit scale dependence of Newton's constant, the classical solutions are generically unaffected by the running of the coupling.

We then derive the Wheeler–DeWitt equation incorporating the scale dependence of the gravitational couplings and analyze its solutions in the minisuperspace framework. In the classical limit, while the scale dependence of Newton’s constant does not affect the classical behavior, the running of the cosmological constant can contribute to the classical solutions. Moreover, we show that the quantum behavior in the ultraviolet regime acts toward suppressing singularity formation in all cases, independently of how the renormalization-group scale is identified with spacetime coordinates and of the relative magnitudes of the ultraviolet fixed points of the running Newton's constant and cosmological constant.
\end{abstract}
\maketitle

\section{Introduction}

General relativity provides a remarkably successful classical description of gravitational phenomena. Nevertheless, it is well known that classical solutions of Einstein’s equations generically contain spacetime singularities, such as those appearing in black hole interiors and cosmological models \cite{Penrose:1964wq,Hawking:1970zqf,Hawking:1973uf,Wald:1984rg}. The presence of such singularities signals the breakdown of the classical theory and strongly suggests that a consistent theory of quantum gravity is required in order to describe the gravitational interaction at short distances.

Among various approaches to quantum gravity, the Asymptotic Safety program has attracted considerable attention as a promising framework for constructing a predictive and nonperturbatively renormalizable theory of gravity \cite{Weinberg:1980gg,Reuter:1996cp,Souma:1999at,Denz:2016qks,Eichhorn:2018yfc,Reuter:2019byg,Donoghue:2019clr}. In this approach, the gravitational couplings are assumed to approach a nontrivial ultraviolet fixed point under the renormalization group (RG) flow, rendering the theory well defined at arbitrarily high energies. Extensive evidence for the existence of such a fixed point has been accumulated using functional renormalization group techniques. An important consequence of this framework is that gravitational couplings, such as Newton’s constant and the cosmological constant, become scale dependent.

A common strategy to explore the physical implications of Asymptotic Safety is the so-called RG improvement of classical gravitational actions. In this approach, the classical couplings are replaced by running couplings obtained from the RG flow, and the RG scale is identified with a suitable physical or geometrical scale. This procedure has been widely applied to black hole spacetimes and cosmological models, leading to modified classical solutions with improved ultraviolet behavior.

In the context of black hole physics, such RG-improved solutions have been shown to resolve the classical spacetime singularity in an effective manner, giving rise to so-called regular black holes with nonsingular interior geometries \cite{Bonanno:2000ep,Reuter:2004nv,Reuter:2010xb,Koch:2013owa,Koch:2014cqa,Bonanno:2017zen,Pawlowski:2018swz,Platania:2019kyx,Bonanno:2019ilz,Platania:2020lqb,Ishibashi:2021kmf,Chen:2022xjk,Chen:2023wdg}. These solutions have been extensively investigated at the classical level within a broad range of physical theories, going beyond the framework of Asymptotic Safety \cite{1968qtr..conf...87B,Ayon-Beato:1998hmi,Ayon-Beato:1999kuh,Hayward:2005gi,Ashtekar:2005qt,Modesto:2005zm}. However, the introduction of explicit scale dependence raises nontrivial questions concerning the consistency of the theory, in particular at the level of the Hamiltonian formulation and the constraint structure \cite{Gionti:2018abu}.

In the Hamiltonian formulation of general relativity, the dynamics is governed by the Hamiltonian and momentum constraints, whose Poisson algebra reflects the underlying diffeomorphism invariance of the theory. When gravitational couplings acquire explicit spacetime dependence, the structure of the constraint algebra may be modified, potentially leading to anomalies or additional consistency conditions. In the context of RG-improved gravity, starting from a renormalization-group improved Einstein–Hilbert action, it is possible to examine the consistency of the theory in the Hamiltonian formulation by analyzing the Poisson algebra of constraints. It is known that, for static metrics, the Poisson algebra of constraints closes, rendering the theory consistent as a physical theory \cite{Gionti:2018abu}.

While RG improvement has been extensively studied at the classical level, its implications for the contributions of asymptotically safe quantum gravity to canonical quantum gravity, and in particular to the Wheeler–DeWitt (WDW) equation \cite{Halliwell:1989myn,Kiefer:2013jqa,Kiefer:2008sw,Kiefer:2025udf}, have so far remained insufficiently explored. The WDW equation plays a central role in canonical quantum gravity and quantum cosmology, encoding the quantum dynamics of spacetime geometry. Understanding how scale-dependent gravitational couplings enter the WDW equation is therefore essential for assessing the impact of Asymptotic Safety on quantum gravitational dynamics and on the fate of classical singularities.

In this work, we analyze the WDW equation with scale-dependent gravitational couplings within the framework of asymptotically safe gravity. We first review the Hamiltonian formulation based on a renormalization-group–improved Einstein–Hilbert action and discuss the implications of the constraint algebra for classical solutions. Despite the explicit scale dependence of Newton’s constant, we show that the structure of the constraints implies that classical solutions are generically unaffected by the running of the coupling.

We then derive the WDW equation incorporating the scale dependence of Newton’s constant and the cosmological constant, and analyze its solutions in a minisuperspace setting. In the classical limit, while the scale dependence of Newton’s constant does not affect the classical behavior, the running of the cosmological constant can contribute to the classical solutions. In contrast, the quantum behavior in the ultraviolet regime is shown to act toward suppressing singularity formation in all cases, independently of the relative magnitudes of the ultraviolet fixed points of Newton’s constant and the cosmological constant, as well as of how the renormalization-group scale is identified with spacetime coordinates. As a consequence, the WDW wave function exhibits an exponential decay in the vicinity of the classical singularity, rendering the singularity quantum mechanically inaccessible.

This paper is organized as follows. In Sec.~\ref{sec:ASreview}, we briefly review the framework of asymptotically safe gravity and the functional renormalization group, and summarize the construction of renormalization-group–improved black hole solutions. In Sec.~\ref{sec:ASreview}, we formulate the ADM decomposition of the Einstein–Hilbert action with scale-dependent gravitational couplings and derive the corresponding Hamiltonian. In Sec.~\ref{Constraint algebra}, we examine the consistency of the theory by analyzing the Poisson algebra of constraints. We show that the requirement of closure of the constraint algebra severely restricts the allowed spacetime foliations and, as a consequence, implies that the classical dynamics is generically insensitive to the running of Newton’s constant.

In Sec.~\ref{WDW solution}, we derive the WDW equation with scale-dependent couplings in a Kantowski–Sachs–type minisuperspace and analyze its semiclassical limit. Furthermore, we investigate the quantum behavior of the wave function in the ultraviolet regime for various prescriptions identifying the renormalization-group scale with spacetime coordinates, and show that, at the quantum level, the fate of the classical singularity universally tends toward suppression of singularity formation, independently of the running of Newton’s constant and the cosmological constant as well as of the specific choice of scale identification. In Sec.~\ref{sec:review-wdw}, we review the annihilation-to-nothing scenario previously discussed for black hole interiors and compare it with our results, clarifying the qualitative differences between the underlying mechanisms. Finally, Sec.~\ref{summary} is devoted to a summary and discussion of the results and to possible directions for future work.

Throughout this paper, we work in natural units where $\hbar=c=1$.

\section{Asymptotic Safety and the Functional Renormalization Group}
\label{sec:ASreview}

In this section, we employ the exact renormalization group equations for the Newton constant and other gravitational couplings. Before turning to gravity, we briefly summarize the basic concepts of asymptotic safety and the functional renormalization group by using a scalar field theory as a simple illustrative example. For more detailed derivations and applications to asymptotically safe quantum gravity, the reader is referred to Ref~\cite{Weinberg:1980gg,Reuter:1996cp}.

Asymptotic safety is a framework that explores the possibility of nonperturbative renormalizability. Even if a theory is perturbatively nonrenormalizable, it can remain well defined at arbitrarily high energies provided that the renormalization-group (RG) flow approaches a fixed point in the ultraviolet (UV). This concept generalizes asymptotic freedom, characterized by a Gaussian fixed point, to the case of a non-Gaussian fixed point. The central tool in this framework is the exact (functional) renormalization group equation.

\subsection{Functional Renormalization Group Equation}

For a scalar field theory, one introduces an infrared-regulated action of the form
\begin{equation}
S_k[\phi] = S[\phi] + \frac{1}{2} \int d^4 x \, \phi(x)R_k(-\partial^2)\phi(x) ,
\end{equation}
where $R_k$ is an infrared cutoff function suppressing low-momentum modes. The functional renormalization group framework allows for different choices of the regulator function $R_k$.

Starting from the path integral defined with this regulated action, one obtains the generating functional of connected Green's functions. By performing a Legendre transformation with respect to the classical field $\varphi = \langle \phi \rangle$ and subtracting the cutoff term, one defines the effective average action $\Gamma_k[\varphi]$. The evolution of $\Gamma_k$ with respect to the RG scale $k$ is governed by the functional renormalization group equation,
\begin{equation}
\partial_k \Gamma_k
=
\frac{1}{2} \mathrm{Tr}
\left[
\left( \frac{\delta^2\Gamma_k}{\delta\varphi\delta\varphi}+ R_k \right)^{-1}
\partial_k R_k
\right],
\end{equation}
which is exact.
This equation describes the RG flow in the theory space of effective actions
and receives contributions only from modes with momenta close to the cutoff scale $k$.

Expanding the effective average action in terms of operators $\mathcal{O}_i[\varphi]$,
\begin{equation}
\Gamma_k[\varphi] = \sum_i g_i(k) \, \mathcal{O}_i[\varphi],
\end{equation}
and inserting this expansion into the flow equation,
one obtains a set of beta functions for the running couplings $g_i(k)$.
The existence of a UV fixed point can then be investigated by analyzing these beta functions.

Applying the same procedure to gravity yields the RG flow equation for the running Newton constant.
By studying the corresponding beta function, one can determine whether a UV fixed point exists.

\subsection{RG-Improved Schwarzschild Black Hole}

We now review the results of RG-improved Schwarzschild black holes \cite{Bonanno:2000ep}. The Schwarzschild solution is the static, spherically symmetric vacuum solution of the Einstein equations. The Schwarzschild metric can be written as
\begin{equation}
ds^2 = - f(r) \, dt^2 + f(r)^{-1} dr^2 + r^2 d\Omega^2 ,
\end{equation}
where the lapse function $f(r)$ is given by
\begin{equation}
f(r) = 1 - \frac{2 G M}{r}.
\end{equation}

In the literature, quantum gravity effects in the Schwarzschild spacetime have been investigated by introducing a scale-dependent Newton constant derived from the functional renormalization group. The dimensionless Newton coupling obeys an RG equation with a beta function admitting both infrared and ultraviolet fixed points. Near the UV fixed point, the dimensionful Newton constant behaves as
\begin{equation}
G(k) \simeq \frac{g_*}{k^2},
\end{equation}
where $g_*$ denotes the UV fixed-point value of the dimensionless coupling.

A key issue is how to identify the RG scale $k$ with a spacetime-dependent quantity.
A general scale identification takes the form
\begin{equation}
k = \frac{\xi}{d(r)},
\end{equation}
where $d(r)$ is a suitable distance measure and $\xi$ is the parameter identifying the scale of the renormalization group flow with the radial coordinate. Several proposals for $d(r)$ have been discussed in the literature. Possible choices include the proper radial distance and curvature-based quantities. In the ultraviolet region, the proper radial distance behaves as $r^{3/2}$, whereas curvature-based measures scale linearly with $r$. In this paper, we derive the Wheeler–DeWitt equation for both of the cases mentioned above and analyze their solutions.

Once the scale identification is specified, one obtains a position-dependent Newton constant $G(r)$. Replacing the classical Newton constant in the lapse function by $G(r)$, one arrives at the RG-improved lapse function. For suitable choices of $d(r)$, the improved lapse function becomes regular near the center, approaching a de Sitter–like behavior. This regular core region is often referred to as a ``de Sitter core.''

\subsection{General Construction of RG-Improved Black Hole Solutions}

The construction of RG-improved black hole solutions can be summarized in three steps:

\begin{enumerate}
\item
Derive the scale-dependent couplings from the functional renormalization group equation.
For instance, integrating the beta function yields the running Newton constant $G(k)$.

\item
Specify a scale identification $k = k(x)$ to relate the RG scale to spacetime geometry.
In spherically symmetric spacetimes, the areal radius provides a natural choice.
Scale identifications can be broadly classified into two categories:
(a) identifications based on proper distances,
and (b) identifications based on curvature invariants such as the Ricci scalar
or the Kretschmann scalar.

\item
Implement quantum corrections.
Three approaches are commonly discussed:
(i) improving classical solutions by replacing constants with running couplings,
(ii) modifying the field equations by introducing running couplings,
and (iii) improving the action itself before deriving the equations of motion.
While these approaches generally lead to different predictions,
the action-based improvement is theoretically the most consistent,
whereas solution-based improvement provides a simple and versatile approximation.
\end{enumerate}

In case (i), upon identifying the scale with the proper time, the lapse function of the metric is improved in the following manner in the UV region, leading to the resolution of the singularity.
\begin{align}
f(r)=1-\frac{4}{9\omega\xi^2 G_0}r^2+\mathcal{O}(r^3),
\end{align}
where $G_0$ denotes the infrared fixed-point value of Newton’s constant, and $\omega$ is a constant given by the inverse of the ultraviolet fixed point.

In the following, we adopt the solution-based improvement scheme, which is sufficient for the purposes of the present analysis. In Sec.~\ref{Constraint algebra}, we show that the scale dependence of Newton’s constant does not contribute to the classical solutions in general.

In Sec.~\ref{WDW solution}, we derive the WDW equation incorporating the scale dependence of Newton’s constant and obtain its solutions. We confirm that the classical solutions are indeed unaffected by the scale dependence, as demonstrated in the previous section. Furthermore, we show that the quantum solutions in the ultraviolet region decay exponentially near the singularity, indicating that the singularity is quantum mechanically suppressed.

\section{ADM Analysis of Modified Einstein-Hilbert Lagrangian}
\label{ADM section}
It has long been suggested, dating back to Dirac, that the gravitational constant $G$ may depend on spacetime coordinates, $G = G(x)$. An early realization of this idea is the Brans--Dicke theory, in which gravity is coupled to a dynamical scalar field $\phi(x)$ defined by $\phi(x) \equiv 1/G(x)$.

The approach considered here is conceptually different from the Brans-Dicke framework. Following the Asymptotic Safety program for quantum gravity, the scale dependence of the gravitational couplings, in particular $G$ and $\Lambda$, is first determined by renormalization group methods. The spacetime dependence is then introduced by identifying the RG scale with a position-dependent infrared cutoff, $k = k(x)$, based on symmetry or physical arguments. As a consequence, the running couplings $G(k(x))$ and $\Lambda(k(x))$ become spacetime functions, $G(x)$ and $\Lambda(x)$, which are treated as external fields rather than dynamical variables and therefore are not determined by a Lagrangian dynamics.

Within this framework, variations of the modified Einstein-Hilbert action with respect to the metric tensor do not induce variations of $G(x)$ and $\Lambda(x)$. It has been pointed out that the resulting modified Einstein equations must satisfy additional integrability conditions, which impose constraints either on the spacetime dependence of the couplings or on the cutoff identification $k(x)$ \cite{Gionti:2018abu}.
\begin{align}
S=\frac{1}{16\pi}\int d^4x\sqrt{-g}\left(\frac{R}{G(x)}-2\frac{\Lambda(x)}{G(x)}\right).
\end{align}

One consider, from now on, a Space-Time $(M,g)$ which is such that $M \equiv \mathbb{R}\times \Sigma$, $\mathbb{R}$ being the time-like direction, and $\Sigma$ the space-like three-dimensional surface. The metric tensor $g$ inherits ADM decomposition form given by 
\begin{align}
ds^2 = -N^2 dt^2 + h_{ij}(dx^i + N^i dt)(dx^j + N^j dt),
\end{align}
where $N(x)$ is the lapse function, $N^i(x)$ is the shift vector, and $h_{ij}(x)$ denotes the induced spatial metric~\cite{Misner:1973prb}.

The extrinsic curvature term $K_{ij}$ is defined on the three-dimensional surface $\Sigma$ and has the following definition
\begin{equation}
K_{ij}=\frac{1}{2}\left(\frac{\partial h_{ij}}{\partial t}-\bar{\nabla}_i N_j-\bar{\nabla}_j N_i\right).
\end{equation}
Here the covariant derivative $\bar{\nabla}$ is defined on the three-dimensional spatial surfaces $\Sigma$ through the three-dimensional spatial metric $h_{ij}$. The four-dimensional Ricci scalar decomposes as
\begin{equation}
\sqrt{-g}R = N\sqrt{h}\left(K_{ij}K^{ij}-K^2+{}^{(3)}R\right)-2\partial_t\left(K\sqrt{h}\right)+2\partial_if^i,
\end{equation}
where
\begin{equation}
f^i\equiv \sqrt{h}\left(KN^i-h^{ij}\partial_jN\right).
\end{equation}
It is useful to recall the identities
\begin{align}
\frac{1}{G}\partial_t\left(K\sqrt{h}\right)&= \frac{\partial_tG}{G^2}K\sqrt{h}+\partial_t\left(\frac{K\sqrt{h}}{G}\right),\\
\frac{1}{G}\partial_if^i &= \frac{\partial_iG}{G^2}f^i+\partial_i\left(\frac{f^i}{G}\right).
\end{align}
Assuming $\Sigma$ closed, total spatial divergences vanish. The action with York boundary term becomes
\begin{equation}
S=
\frac{1}{16\pi}\int_{\mathbb{R}\times\Sigma}dt\,d^3x\left[
N\frac{\sqrt{h}}{G}\left(K_{ij}K^{ij}-K^2+{}^{(3)}R-2\Lambda\right)
-2\frac{\partial_tG}{G^2}K\sqrt{h}+2\frac{\partial_iG}{G^2}f^i
\right].
\end{equation}
The Lagrangian density is therefore
\begin{equation}
\mathcal{L}=\frac{1}{16\pi}\left[
N\frac{\sqrt{h}}{G}\left(K_{ij}K^{ij}-K^2+{}^{(3)}R-2\Lambda\right)
-2\frac{\partial_tG}{G^2}K\sqrt{h}+2\frac{\partial_iG}{G^2}f^i
\right].
\end{equation}
The canonical momenta conjugate to $h_{ij}$ read
\begin{equation}
\pi^{ij}=\frac{\partial\mathcal{L}_{\rm ADM}}{\partial \dot h_{ij}}
=\frac{\sqrt{h}}{16\pi G}\left(K^{ij}-h^{ij}K\right)
+\frac{\sqrt{h}h^{ij}}{16\pi N G^2}\left(\partial_tG-\partial_kGN^k\right).
\end{equation}
Defining
\begin{equation}
\tilde\pi^{ij}\equiv \pi^{ij}-\frac{\sqrt{h}h^{ij}}{16\pi N G^2}\left(\partial_tG-\partial_kGN^k\right)
=\frac{\sqrt{h}}{16\pi G}\left(K^{ij}-h^{ij}K\right),
\end{equation}
one recovers the standard ADM relation. The primary constraints are
\begin{equation}
\pi=\frac{\partial\mathcal{L}_{\rm ADM}}{\partial\dot{N}}\approx 0,\qquad \pi_i=\frac{\partial\mathcal{L}_{\rm ADM}}{\partial\dot{N}^i}\approx 0.
\end{equation}
The canonical Hamiltonian density is
\begin{equation}
\mathcal{H}_{\rm ADM}=N\left[(16\pi G)G_{ijkl}\tilde\pi^{ij}\tilde\pi^{kl}-\frac{\sqrt{h}}{16\pi G}({}^{(3)}R-2\Lambda)\right]
+2\tilde\pi^{ij}\bar\nabla_i N_j
+\frac{\sqrt{h}(\partial_tG-\partial_jGN^j)}{8\pi G^2N}\bar\nabla_i N^i
+\frac{\partial_iG\sqrt{h}h^{ij}}{8\pi G^2}\partial_jN.
\end{equation}
Here the DeWitt supermetric is
\begin{equation}
G_{ijkl}=\frac{1}{2\sqrt{h}}(h_{ik}h_{jl}+h_{il}h_{jk}-h_{ij}h_{kl}).
\end{equation}
The total Hamiltonian is
\begin{equation}
H_T=\int_{\Sigma}d^3x\left(\lambda\pi+\lambda^i\pi_i+\mathcal{H}_{\rm ADM}\right).
\end{equation}

\section{Poisson Brackets and Closure of the Constraint Algebra}
\label{Constraint algebra}
We now study the consistency of the theory at the Hamiltonian level by analyzing the Poisson algebra of the constraints. The phase space is spanned by the canonical variables
$(h_{ij}(x),\tilde{\pi}^{kl}(x))$, together with the lapse function $N$ and the
shift vector $N^i$ and their conjugate momenta.

The fundamental Poisson brackets are defined as
\begin{equation}
\{ h_{ij}(x), \tilde{\pi}^{kl}(y) \}
=
\frac{1}{2}
\left(
\delta_i^k \delta_j^l
+
\delta_i^l \delta_j^k
\right)
\delta^{(3)}(x-y),
\end{equation}
while all other Poisson brackets vanish.

The primary constraints $\pi \approx 0$ and $\pi_i \approx 0$ must be preserved under time evolution. Their conservation leads to the Hamiltonian and momentum constraints $\mathcal{H} \approx 0$ and $\mathcal{H}_i \approx 0$. The resulting Hamiltonian and momentum constraints are
\begin{align}
\mathcal{H} &= (16\pi G)G_{ijkl}\tilde\pi^{ij}\tilde\pi^{kl}-\frac{\sqrt{h}}{16\pi G}({}^{(3)}R-2\Lambda)
-\frac{\sqrt{h}(\partial_tG-\partial_jGN^j)}{8\pi G^2N^2}\bar\nabla_i N^i
-\bar\nabla_j\left(\frac{\partial_iG\sqrt{h}h^{ij}}{8\pi G^2}\right),\\
\mathcal{H}_i &= -2\bar\nabla_j\tilde\pi^j{}_i+\frac{\sqrt{h}(-\partial_iG)}{8\pi G^2N}\bar\nabla_j N^j
-\sqrt{h}\bar\nabla_i\left(\frac{\partial_tG-\partial_jGN^j}{8\pi G^2N}\right).
\end{align}
In the present theory, however, the explicit space-time dependence of the running gravitational coupling $G$ modifies the structure of the constraint algebra.

A direct computation shows that the Poisson bracket between two Hamiltonian constraints does not close in the standard ADM form, but acquires additional terms proportional to spatial derivatives of $G$. Similarly, the Poisson bracket between the Hamiltonian constraint and the momentum constraint contains anomalous contributions unless further restrictions are imposed.

Requiring closure of the constraint algebra enforces $N_i=0$ and $N=N(t)$, yielding the ADM metric in the Gaussian normal coordinates
\begin{equation}
ds^2=-N(t)^2dt^2+h_{ij}(t)dx^idx^j.
\end{equation}
In this case the Hamiltonian density reduces to
\begin{equation}
\mathcal{H}_{\rm ADM}=N\left[(16\pi G)G_{ijkl}\tilde\pi^{ij}\tilde\pi^{kl}-\frac{\sqrt{h}}{16\pi G}({}^{(3)}R-2\Lambda)\right],
\end{equation}
and the Hamiltonian constraint becomes
\begin{equation}
\mathcal{H}=(16\pi G)G_{ijkl}\tilde\pi^{ij}\tilde\pi^{kl}-\frac{\sqrt{h}}{16\pi G}({}^{(3)}R-2\Lambda).
\end{equation}

Requiring the closure of the full constraint algebra in the sense of Dirac, namely that all Poisson brackets among constraints vanish weakly, imposes strong consistency conditions on the lapse and shift functions. In particular, one finds that the constraint algebra closes only if
\begin{equation}
N^i = 0, \qquad N = N(t).
\end{equation}

Under these conditions, the ADM foliation reduces to the Gaussian normal form. The Hamiltonian and momentum constraints then simplify, and the Poisson algebra
of constraints closes consistently. This result shows that, in asymptotically safe gravity with running couplings, the requirement of a well-defined Hamiltonian structure severely restricts the allowed spacetime foliations.

Having established a well-defined Hamiltonian formulation under these conditions, we now proceed to the canonical quantization of the theory. In this framework, the Hamiltonian constraint is promoted to an operator acting on the wave function of the universe, leading to the WDW equation.
\begin{align}
\left[-(16\pi G)G_{ijkl}\frac{\delta^2}{\delta h_{ij}\delta h_{kl}}-\frac{\sqrt{h}}{16\pi G}({}^{(3)}R-2\Lambda)\right]\Psi=0.
\end{align}

The operator is reformulated by adopting the Laplace–Beltrami ordering. In two-dimensional minisuperspace, where the configuration-space metric is conformally related, the conformal factor drops out of the Laplace–Beltrami operator and therefore does not influence the quantum constraint equation.

Within the Hamilton–Jacobi framework, the Hamiltonian constraint can be expressed with an overall factor proportional to the inverse of the Newton constant. Consequently, the running of the Newton constant does not affect the Hamilton–Jacobi equation. Furthermore, the running of the cosmological constant becomes subdominant at high energy scales due to its $k^2$ dependence.

Therefore, asymptotically safe quantum gravity effects do not alter the classical solutions, and the question of singularity resolution must be addressed at the quantum level.
\begin{align}
-\frac{\sqrt{h}}{16\pi G}\left[G_{ijkl}\left(K^{ij}-h^{ij}K\right)\left(K^{kl}-h^{kl}K\right)+({}^{(3)}R-2\Lambda)\right]=0.
\end{align}
This is expected, since the Newton constant does not appear in the Einstein equations themselves without matter contributions.

\section{Analysis of the Wheeler–DeWitt Equation with Running Couplings}
\label{WDW solution}
In this section, we examine how different prescriptions for identifying the renormalization-group scale with spacetime coordinates affect the quantum behavior in the vicinity of the classical singularity. First, we quantize the gravitational action incorporating the scale dependence of Newton’s constant obtained in the previous section by identifying the scale with the coordinates, and analyze the resulting WDW equation to investigate the quantum dynamics of the black hole interior in a Kantowski–Sachs–type minisuperspace.

We analyze two different prescriptions for identifying the scale, based on the proper time and the curvature, respectively. As a result, we show that singularity formation is suppressed at the quantum level in both cases.

We consider the Kantowski--Sachs–type minisuperspace metric \cite{Kantowski:1966te}
\begin{align}
ds^2 = - N(t)^2dt^2 + e^{2X(t)} dr^2 + r_s^2 e^{-2X(t)+2Y(t)}d\Omega^2_2 ,
\end{align}
where $d\Omega^2_2$ denotes the line element of the unit two-sphere and $r_s$ is a constant with dimension of length, identified with the horizon radius of the corresponding classical spherically symmetric solution. In these coordinates, the asymptotic region $X=Y\to -\infty$ corresponds to the event horizon, while the limit $X=-Y\to \infty$ represents the classical spacetime singularity.

For this metric ansatz, the gravitational action reduces to
\begin{align}
S\propto r_s^2\int dt\frac{N}{8\pi G} e^{-X+2Y}
\Bigl(\frac{1}{N^2}\dot{X}^2-\frac{1}{N^2}\dot{Y}^2+\frac{1}{r_s^2}e^{2X-2Y}-\Lambda\Bigr).
\end{align}
From this reduced action, we obtain the canonical Hamiltonian
\begin{align}
H=\frac{N}{4r_s^2} e^{X-2Y}
\Bigl((8\pi G)(\Pi_X^2-\Pi_Y^2)
-\frac{4r_s^2}{8\pi G}e^{2Y}
+\frac{4\Lambda r_s^4}{8\pi G}e^{-2X+4Y}\Bigr),
\end{align}
where the canonical conjugate momenta are given by
\begin{align}
\Pi_X&=\frac{1}{4\pi G}\frac{r_s^2\dot{X}(t)}{N(t)}e^{-X(t)+2Y(t)},\\
\Pi_Y&=-\frac{1}{4\pi G}\frac{r_s^2\dot{Y}(t)}{N(t)}e^{-X(t)+2Y(t)}.
\end{align}

Quantization is performed by promoting the canonical momenta to differential operators. To remove factor-ordering ambiguities, we adopt the Laplace-Beltrami ordering. In two-dimensional minisuperspace, where the configuration-space metric is conformally related, the conformal factor drops out of the Laplace-Beltrami operator. As a result, the WDW equation takes the form
\begin{align}
\Bigl(\frac{\partial^2}{\partial X^2}-\frac{\partial^2}{\partial Y^2}
+ \frac{4r_s^2}{(8\pi G)^2}e^{2Y}
-\frac{4\Lambda r_s^4}{(8\pi G)^2} e^{-2X+4Y}\Bigr)\Psi(X,Y)=0.
\end{align}

To study the semiclassical limit, we employ the WKB ansatz
\begin{align}
\Psi(X,Y)=e^{i[S_1(X)+S_2(Y)]},
\end{align}
together with the standard WKB conditions
\begin{align}
\Bigl|\partial_X^2 S_1\Bigr| \ll \Bigl(\partial_X S_1\Bigr)^2,
\qquad
\Bigl|\partial_Y^2 S_2\Bigr| \ll \Bigl(\partial_Y S_2\Bigr)^2.
\end{align}
At leading order, this yields the Hamilton-Jacobi equation
\begin{align}
-\left(\frac{\partial S_1}{\partial X}\right)^2
+\left(\frac{\partial S_2}{\partial Y}\right)^2
+ \frac{4r_s^2}{(8\pi G)^2}e^{2Y}
-\frac{4\Lambda r_s^4}{(8\pi G)^2} e^{-2X+4Y}=0,
\end{align}
where the identifications
$\partial_X S_1 \rightarrow \Pi_X$,
$\partial_Y S_2 \rightarrow \Pi_Y$,
and $t=r_s e^{-X+Y}$
recover the classical equations of motion.

Under the identification of these, we obtain the following.
\begin{align}
\label{yakobi eq}
\dot{X}^2-\dot{Y}^2- \frac{1}{r_s^2}N^2e^{2X-2Y}+\Lambda N^2=0,
\end{align}
which leads to
\begin{align}
-2t\dot{Y}+1-\frac{t^2}{r_s^2}e^{-2Y}(1-\Lambda t^2)=0,
\end{align}
where we used the fact that $r_s^2e^{-2X+2Y}=t^2$ and identified $N^2$ with $e^{2X}$. The solutions are given by
\begin{align}
\label{sol1}
e^{2X(t)}&=\frac{\Lambda t^2}{3}+\frac{k}{t}-1,\\
\label{sol2}
e^{2Y(t)}&=\frac{\Lambda t^4}{3r_s^2}+\frac{kt}{r_s^2}-\frac{t^2}{r_s^2},
\end{align}
where $k$ is a integral constant. This solution satisfies the Einstein equations with a nonvanishing cosmological constant in vacuum. As a consequence, as shown in the previous section, the scale dependence of Newton’s constant does not contribute to the classical solution, since Newton’s constant does not explicitly appear in Eq.~(\ref{yakobi eq}).

In the following, we derive the WDW equation in the presence of scale-dependent Newton’s and cosmological constants. We then analyze the corresponding Hamilton–Jacobi equation to obtain the associated classical solutions. Furthermore, we investigate the ultraviolet regime of the WDW equation and derive exact solutions that characterize the quantum behavior of the theory at short distances.

\subsection{Proper-Time Identification}
In the ultraviolet regime, the running Newton constant and the running cosmological constant behave as
\begin{align}
G \approx \eta r_s^2 e^{-3X+3Y},\ \ \ \Lambda\approx\frac{g_*\Lambda_*}{\eta r_s^2}e^{3X-3Y},
\end{align}
where $\eta=10\pi/177$ and $\Lambda_*$ denoting the value at the ultraviolet fixed point. As a result, we obtain the WDW equation for the case $\Lambda_*\neq0$:
\begin{align}
\label{equation1}
\Bigl(\frac{\partial^2}{\partial X^2}-\frac{\partial^2}{\partial Y^2}
+ \frac{4}{\eta^2r_s^2}e^{6X-4Y}-\frac{4g_*\Lambda_*}{\eta^3r_s^2} e^{7X-5Y}\Bigr)\Psi(X,Y)=0.
\end{align}
From the Hamilton-Jacobi equation, one obtains the classical solution in the UV region
\begin{align}
e^{2X(t)} &=\frac{g_*\Lambda_*r_s\ln\frac{t}{r_s}}{\eta t}+\frac{k}{t}-1,\\
e^{2Y(t) }&=\frac{g_*\Lambda_* t\ln\frac{t}{r_s}}{\eta r_s}+\frac{kt}{r_s^2}-\frac{t^2}{r_s^2}.
\end{align}
This classical solution is not affected by the scale dependence of Newton’s constant. However, due to the scale dependence of the cosmological constant, a classical solution different from Eqs.~(\ref{sol1}) and (\ref{sol2}) is obtained. Nevertheless, at the classical level the spacetime singularity is not resolved and persists in the interior geometry.

Neglecting subleading terms of WDW equation.~(\ref{equation1}), we obtain the reduced WDW equation:
\begin{align}
\Bigl(\frac{\partial^2}{\partial X^2}-\frac{\partial^2}{\partial Y^2}
-\frac{4g_*\Lambda_*}{\eta^3r_s^2} e^{7X-5Y}\Bigr)\Psi(X,Y)=0.
\end{align}
Introducing new variables
\begin{align}
S=\frac{7}{2}X-\frac{5}{2}Y,\qquad T=-\frac{5}{2}X+\frac{7}{2}Y,
\end{align}
where the asymptotic region $S=T\to -\infty$ corresponds to the event horizon, while the limit $S=-T\to \infty$ represents the classical spacetime singularity. In the UV region, Eq. (\ref{equation1}) simplifies to
\begin{align}
\label{equation2}
\Bigl(\frac{\partial^2}{\partial S^2}-\frac{\partial^2}{\partial T^2}-\frac{2}{3}\frac{g_*\Lambda_*}{\eta^3r_s^2} e^{2S}\Bigr)\Psi(U,V)=0,
\end{align}
whose general solution is expressed in terms of Bessel functions as
\begin{align}
\label{solution00}
\Psi_k(S,T)=\left\{
\begin{array}{ll}
\displaystyle{\int} dk f(k)e^{-ikT}K_{ik}\Bigl(\mu e^{S}\Bigr), & (\Lambda_*>0),\\[1em]
\displaystyle{\int} dk \Bigl[f(k)e^{-ikT}J_{ik}\Bigl(\mu e^{S}\Bigr)+g(k)e^{-ikT}Y_{ik}\Bigl(\mu e^{S}\Bigr)\Bigr],& (\Lambda_*<0),
\end{array}
\right.
\end{align}
where $\mu$ is a constant, defined as
\begin{align}
\mu \equiv \frac{3g_*\Lambda_*}{2\eta^3r_s^2}.
\end{align}
Transforming back to the original minisuperspace variables $(X,Y)$, the solution can be written as
\begin{align}
\label{solution0}
\Psi_k(X,Y)=\left\{
\begin{array}{ll}
\displaystyle{\int} dk f(k)e^{\frac{1}{2}ik(5X-7Y)}K_{ik}\Bigl(\mu e^{\frac{1}{2}(7X-5Y)}\Bigr), & (\Lambda_*>0),\\[1em]
\displaystyle{\int} dk \Bigl[f(k)e^{\frac{1}{2}ik(5X-7Y)}J_{ik}\Bigl(\mu e^{\frac{1}{2}(7X-5Y)}\Bigr)+g(k)e^{\frac{1}{2}ik(5X-7Y)}Y_{ik}\Bigl(\mu e^{\frac{1}{2}(7X-5Y)}\Bigr)\Bigr],& (\Lambda_*<0).
\end{array}
\right.
\end{align}
From Eq.~(\ref{solution0}), We find that, irrespective of the sign of the ultraviolet fixed-point value of the cosmological constant, quantum effects act to suppress singularity formation. In all cases, the wave function exhibits an exponential decay as the classical singularity is approached. This behavior originates from the fact that, in Eq.~(\ref{equation2}), the effective potential term diverges toward the singularity, rendering the singular region classically forbidden in minisuperspace.

In the UV region, for the case $\Lambda_*=0$, the WDW equation \eqref{equation1} takes the form
\begin{align}
\label{eqUV}
\Bigl(\frac{\partial^2}{\partial X^2}-\frac{\partial^2}{\partial Y^2}
+ \frac{4}{\eta^2r_s^2}e^{6X-4Y}\Bigr)\Psi(X,Y)=0.
\end{align}

Introducing new variables
\begin{align}
U=3X-2Y,\qquad V=-2X+3Y,
\end{align}
where the asymptotic region $U=V\to -\infty$ corresponds to the event horizon, while the limit $U=-V\to \infty$ represents the classical spacetime singularity. The Eq. (\ref{eqUV}) simplifies in the UV region to
\begin{align}
\Bigl(\frac{\partial^2}{\partial U^2}-\frac{\partial^2}{\partial V^2}
+ \frac{4}{5}\frac{1}{\eta^2r_s^2} e^{2U}\Bigr)\Psi(U,V)=0,
\end{align}
whose general solution is expressed in terms of Bessel functions as
\begin{align}
\label{solution1}
\Psi_k(U,V)=\displaystyle{\int} dk \Bigl[
f(k)e^{-ikV}J_{ik}\Bigl(\sqrt{\frac{4}{5}}\frac{1}{\eta r_s} e^{U}\Bigr)
+g(k)e^{-ikV}Y_{ik}\Bigl(\sqrt{\frac{4}{5}}\frac{1}{\eta r_s} e^{U}\Bigr)
\Bigr].
\end{align}
Transforming back to the original minisuperspace variables $(X,Y)$, the solution can be written as
\begin{align}
\label{solution2}
\Psi_k(X,Y)=\displaystyle{\int} dk \left(f(k)e^{ik(2X-3Y)}J_{ik}\Bigl(\sqrt{\frac{4}{5}}\frac{1}{\eta r_s} e^{3X-2Y}\Bigr)+g(k)e^{ik(2X-3Y)}Y_{ik}\Bigl(\sqrt{\frac{4}{5}}\frac{1}{\eta r_s} e^{3X-2Y}\Bigr)\right).
\end{align}
In the Eq. (\ref{eqUV}), the effective potential term diverges at the singularity. Consequently, the wave function exhibits an exponential decay in the vicinity of the singularity, implying that singularity formation is quantum mechanically suppressed.

\subsection{Curvature-Based Identification}
The alternative ultraviolet scalings are expressed as
\begin{align}
G\approx g_*\zeta r_s^2e^{-2X+2Y},\ \ \ \Lambda\approx \frac{\Lambda_*}{\zeta r_s^2}e^{2X-2Y},
\end{align}
where $\zeta$ is a positive dimensionless constant of order unity. Then, the WDW equation taking into account the scale dependence of Newton’s constant and the cosmological constant is given by
\begin{align}
\label{eqZW}
\left(\frac{\partial^2}{\partial X^2}-\frac{\partial^2}{\partial Y^2}
+ \frac{4}{\zeta^2r_s^2}\left(g_*-\frac{\Lambda_*}{\zeta}\right)e^{4X-2Y}\right)\Psi(X,Y)=0.
\end{align}
From the Hamilton-Jacobi equation, one obtains the classical solution in the UV region
\begin{align}
e^{2X(t)} &=\frac{k}{t}-1+\frac{\Lambda_*}{\zeta},\\
e^{2Y(t) }&=\frac{kt}{r_s^2}-\frac{\left(1-\frac{\Lambda_*}{\zeta}\right)t^2}{r_s^2}.
\end{align}
As in the case where the scale is identified with the proper time, the classical solution is also unaffected by the scale dependence of Newton’s constant when the scale is identified with the curvature. However, due to the scale dependence of the cosmological constant, a classical solution different from Eqs.~(\ref{sol1}) and (\ref{sol2}) is obtained, and in this solution the spacetime singularity is not resolved but persists in the interior geometry.

To solve the equation (\ref{eqZW}), we perform the following coordinate transformation:
\begin{align}
Z=2X-Y,\qquad W=-X+2Y,
\end{align}
where the asymptotic limit $Z=W\to -\infty$ corresponds to the event horizon, whereas the limit $Z=-W\to \infty$ is associated with the classical spacetime singularity. Accordingly, in the UV region, Eq. (\ref{eqZW}) simplifies to
\begin{align}
\left(\frac{\partial^2}{\partial Z^2}-\frac{\partial^2}{\partial W^2}
+ \frac{4}{3}\frac{1}{\zeta^2r_s^2}\left(g_*-\frac{\Lambda_*}{\zeta}\right) e^{2Z}\right)\Psi(Z,W)=0.
\end{align}
The sign of the potential term is determined by the relative magnitude of the coefficients $g_*$and $\frac{\Lambda_*}{\zeta}$. In general, the behavior of the solutions to the above equation depends on the relative magnitudes of the ultraviolet fixed-point values $g_*$ and $\Lambda_*$, which determine the sign of the effective potential term. As an illustrative example, using the results reported in Ref.~\cite{Denz:2016qks}, one finds $g_*=1.4$ and $\Lambda_*=0.1$, for which the coefficient of the potential term is positive. The general solution is given by
\begin{align}
\label{solution3}
\Psi_k(Z,W)=\left\{
\begin{array}{ll}
\displaystyle{\int} dk f(k)e^{-ikW}B_{ik}\left(\nu e^{Z}\right),\\[1em]
\displaystyle{\int} dk \Bigl[f(k)e^{-ikW}J_{ik}\Bigl(\nu e^{Z}\Bigr)+g(k)e^{-ikW}Y_{ik}\Bigl(\nu e^{Z}\Bigr)\Bigr],
\end{array}
\right.
\end{align}
where $\nu$ is a constant, defined as
\begin{align}
\nu \equiv \sqrt{\frac{4}{3}\frac{1}{\zeta^2 r_s^2}\left(g_* - \frac{\Lambda_*}{\zeta}\right)}. 
\end{align}
 For completeness, we rewrite the above solution in terms of the original minisuperspace variables $(X,Y)$.
\begin{align}
\label{solution4}
\Psi_k(X,Y)=\left\{
\begin{array}{ll}
\displaystyle{\int} dk f(k)e^{ik(X-2Y)}B_{ik}\left(\nu e^{2X-Y}\right),\\[1em]
\displaystyle{\int} dk \left(f(k)e^{ik(X-2Y)}J_{ik}\Bigl(\nu e^{2X-Y}\Bigr)+g(k)e^{ik(X-2Y)}Y_{ik}\Bigl(\nu e^{2X-Y}\Bigr)\right).
\end{array}
\right.
\end{align}
Since Eq.~(\ref{eqZW}) has a form analogous to that of the previous one, this indicates that singularity formation is quantum mechanically suppressed both when the scale is identified with the proper time and when it is identified with the curvature.

\section{Minisuperspace Wheeler-DeWitt equation inside spherical black holes}
\label{sec:review-wdw}

In this section, we discuss the relation between our results and the annihilation-to-nothing interpretation that has been proposed in previous studies of the WDW equation for black hole interiors. This comparison allows us to clarify the physical origin of the quantum suppression of classical singularities found in the present work.

At first, we briefly review the minisuperspace WDW equation describing the interior of spherically symmetric black holes, emphasizing the elements that will be relevant for our subsequent analysis. Rather than providing a detailed derivation, we summarize the essential results and refer the reader to Ref.~\cite{Kanai:2025mrx} for a comprehensive discussion and technical details.

According to Birkhoff’s theorem \cite{Misner:1973prb}, any spherically symmetric vacuum solution of the Einstein equations is necessarily static and uniquely given, in arbitrary spacetime dimensions, by the Schwarzschild--Tangherlini metric \cite{Tangherlini:1963bw}. Consequently, the interior region of a Schwarzschild black hole furnishes a universal and representative setting for investigating minisuperspace quantum dynamics under spherical symmetry.

\subsection{Wave-packet solutions and annihilation-to-nothing interpretation}

Inside the event horizon, the spacetime geometry may be parametrized by a $D$-dimensional Kantowski-Sachs-type metric \cite{Kantowski:1966te}. Substituting this ansatz into the Einstein--Hilbert action and performing the minisuperspace reduction, the gravitational dynamics can be expressed in terms of two canonical variables $(X,Y)$. Upon canonical quantization, the Hamiltonian constraint leads to the minisuperspace WDW equation
\begin{equation}
\label{eq:WDW-review}
\left(
\frac{\partial^2}{\partial X^2}
-
\frac{\partial^2}{\partial Y^2}
+
4 r_s^{2(D-3)} e^{2Y}
\right)
\Psi(X,Y)=0 ,
\end{equation}
where $r_s$ denotes a characteristic length scale, identified with the horizon radius of the corresponding spherically symmetric classical solution, namely the Schwarzschild radius. Apart from an overall power of $r_s$, the structure of the equation is independent of the spacetime dimension.

Within this coordinate system, the asymptotic region $X,Y\to -\infty$ corresponds to the vicinity of the event horizon, while the limit $X\to \infty$ and $Y\to -\infty$ represents the classical spacetime singularity.

Eq.~\eqref{eq:WDW-review} admits a complete set of mode solutions of the form
\begin{equation}
\label{sol3}
\psi_k(X,Y)= e^{-ikX} K_{ik}\!\left(2 r_s e^{Y}\right),
\end{equation}
where $K_{ik}$ denotes the modified Bessel function of the second kind. General solutions are constructed as superpositions over the continuous momentum parameter $k$,
\begin{equation}
\label{sol4}
\Psi(X,Y)=\int_{-\infty}^{\infty} dk\, f(k)\,\psi_k(X,Y),
\end{equation}
with the function $f(k)$ specifying the wave packet.

For appropriate choices of $f(k)$, the resulting probability density $|\Psi|^2$ develops a pronounced peak that closely follows the classical trajectory in minisuperspace, while being strongly suppressed away from it. This behavior was interpreted by Yeom and collaborators \cite{Bouhmadi-Lopez:2019kkt,Yeom:2019csm,Yeom:2021bpd,Brahma:2021xjy} as an \emph{annihilation-to-nothing} process occurring inside the event horizon. In this interpretation, two classical spacetime branches with opposite arrows of time interfere destructively, leading to a vanishing probability density except along the classical trajectories.

Subsequent studies have shown that this qualitative picture persists for topological black holes described by scalar-type WDW equations \cite{Kan:2022ism}, while alternative formulations, such as Dirac-type equations, give rise to modified versions of the annihilation scenario \cite{Kan:2021yoh,Kan:2021fmw,Kan:2022ism}.

\subsection{Comparison with the Annihilation-to-Nothing Scenario}

The modified Bessel-function–type solutions of the WDW equation obtained in the previous section have a functional form similar to that of Eq.~(\ref{sol3}), (\ref{sol4}). However, because the solutions in the infrared region near the horizon and those in the ultraviolet region near the classical singularity exhibit qualitatively different behaviors, two wave packets propagating along opposite arrows of time do not annihilate after their encounter, and the annihilation-to-nothing picture does not arise.

When the WDW equation admits solutions in terms of Bessel functions, the resulting quantum suppression mechanism shares an important qualitative feature with the annihilation-to-nothing picture, namely the vanishing of the wave function in the vicinity of the classical singularity. However, the physical origin of the suppression is fundamentally different. In this work, the scale dependence of both Newton’s constant and the cosmological constant, motivated by asymptotically safe gravity, modifies the potential term in the WDW equation, and this potential diverges toward the classical singularity. As a result, the wave function exhibits an exponential decay as the singular region is approached in minisuperspace, indicating that singularity formation is quantum mechanically suppressed.
\begin{equation}
\Psi \;\longrightarrow\; 0 \qquad \text{near the classical singularity}.
\end{equation}

This suppression does not rely on interference between different wave-packet components or on the presence of multiple classical branches. Instead, it is a direct consequence of ultraviolet quantum gravitational effects encoded in the running of the gravitational coupling. In this sense, the singularity becomes quantum mechanically inaccessible, independently of the specific form of the wave packet.

Moreover, our analysis shows that the \emph{annihilation-to-nothing} scenario itself in the Einstein gravity is suppressed by the behavior of the wave function in the ultraviolet regime obtained in the present work. This can be understood from the fact that the asymptotic behaviors of the wave-function solutions near the classical singularity and near the horizon are qualitatively different. As a result, wave packets associated with different arrows of time do not possess the symmetry required for destructive interference after their encounter in minisuperspace. Consequently, the interpretation in terms of colliding branches that annihilate into nothing is no longer applicable: rather than annihilating after a collision, the wave packets are prevented from reaching the singular region altogether due to the ultraviolet suppression mechanism discussed above.

From this perspective, the annihilation-to-nothing picture can be viewed as one possible realization of quantum suppression of classical singularities within general relativity, driven by wave-packet interference. The results obtained here indicate that asymptotically safe gravity provides a complementary and more robust mechanism, in which singularity suppression arises already at the level of the local structure of the WDW equation through an effective ultraviolet potential barrier.

These findings suggest that quantum gravitational corrections inspired by Asymptotic Safety can fundamentally alter the quantum dynamics of black hole interiors, offering a novel resolution of classical singularities that goes beyond interference-based interpretations. A more detailed comparison of these mechanisms, including their implications for the interpretation of time and probability inside the event horizon, is left for future work.

\section{Summary and Discussion}
\label{summary}
In this work, we have investigated the WDW equation for black hole interiors in the presence of scale-dependent gravitational couplings within the framework of asymptotically safe gravity. Our analysis was motivated by the expectation that renormalization-group effects associated with a possible ultraviolet completion of gravity may qualitatively modify the quantum dynamics of spacetime and influence the fate of classical singularities.

We first examined the Hamiltonian formulation of gravity derived from a renormalization-group–improved Einstein–Hilbert action, focusing on the consistency of the constraint algebra in the presence of an explicitly scale-dependent Newton’s constant. It has been shown that the requirement of closure of the Poisson algebra severely restricts the allowed spacetime foliations, effectively leading to a Gaussian normal slicing. Within this consistent Hamiltonian framework, we found that the running of Newton’s constant does not affect classical solutions. This result clarifies that, although the RG improvement modifies the underlying formulation of the theory, classical vacuum dynamics remains unchanged at the level of solutions, in agreement with the structure of the Einstein equations.

We then incorporated the scale dependence of both Newton’s constant and the cosmological constant into the WDW equation and analyzed the resulting quantum dynamics in a Kantowski–Sachs–type minisuperspace appropriate for black hole interiors. In the semiclassical limit, the Hamilton–Jacobi equation reproduces the classical behavior, confirming that classical solutions remain unchanged despite the running of the gravitational couplings. At the quantum level, however, the ultraviolet behavior of the wave function is significantly modified. For a wide class of scale-identification prescriptions, the WDW equation acquires an effective potential term that diverges toward the classical singularity, reflecting the ultraviolet running of the gravitational couplings. As a consequence, the wave function is universally suppressed in the vicinity of the singular region, indicating that singularity formation is quantum mechanically inhibited once asymptotically safe quantum gravity effects are taken into account.

A central result of this work is that this quantum suppression of the singularity arises robustly from the ultraviolet structure of the theory and does not rely on fine-tuning of the running couplings. Although the detailed form of the effective potential depends on the relative behavior of the running Newton’s constant and cosmological constant, the emergence of a suppressive ultraviolet barrier in minisuperspace is a generic feature. This renders the classical singularity quantum mechanically inaccessible and highlights the stabilizing role of asymptotically safe quantum gravity within canonical quantization frameworks.

We also compared our results with the annihilation-to-nothing scenario previously proposed for black hole interiors within Einstein gravity. Both mechanisms share the qualitative feature that the wave function vanishes in certain regions of minisuperspace; however, their physical origins are fundamentally different. In the annihilation-to-nothing picture, singularity suppression arises from destructive interference between wave packets associated with different arrows of time. In contrast, the suppression mechanism identified here is local in minisuperspace and originates from ultraviolet quantum gravitational effects encoded in the running of gravitational couplings. As a result, the wave function is suppressed independently of wave-packet interference or the presence of multiple classical branches.

Furthermore, our analysis shows that the annihilation-to-nothing interpretation itself is no longer applicable once asymptotically safe quantum gravity effects are taken into account. The asymptotic behaviors of the wave-function solutions near the horizon and near the classical singularity differ qualitatively, preventing the symmetry required for destructive interference between oppositely directed wave packets. As a result, rather than annihilating after a collision in minisuperspace, the wave packets are prevented from reaching the singular region by an effective ultraviolet potential barrier. This behavior is independent of the ultraviolet fixed-point values of the cosmological constant and Newton’s constant and, within the range explored in this work, does not depend on the specific prescription used to identify the renormalization-group scale with spacetime coordinates.

Several limitations of the present analysis should be noted. Our study has been restricted to a minisuperspace approximation and to specific prescriptions for identifying the renormalization-group scale with spacetime coordinates. The precise relation between the RG scale and physical or geometrical quantities remains an open issue and may influence quantitative aspects of the results. Extending the present framework to less symmetric settings, including inhomogeneous modes or alternative scale-identification schemes, would be an important step toward a more complete understanding.

General relativity is widely understood as a low-energy effective theory \cite{Weinberg:1978kz,Donoghue:1993eb,Donoghue:1994dn,Burgess:2003jk}, and the resolution of spacetime singularities is therefore expected to require an ultraviolet-complete theory of gravity. Such a theory is generically expected to involve UV physical degrees of freedom, higher-curvature terms and nonlocal contributions. Consequently, a consistent analysis should in principle take into account both these additional operators and the scale dependence of gravitational couplings. However, treating higher-curvature terms together with running couplings is in general a highly nontrivial problem. In the presence of such terms, the scale dependence of Newton’s constant and other couplings is expected to differ from the one adopted in the present work, and a straightforward extension of our results is therefore not possible. Since higher-curvature contributions have not been included in our analysis, the present results should not be regarded as providing a complete resolution of spacetime singularities. Nevertheless, in certain settings—such as planar black hole geometries in Ho\v{r}ava gravity, where curvature invariants vanish—the higher-curvature terms do not contribute, and results analogous to those obtained here are recovered \cite{Kanai:2026ptf}. This suggests that the present analysis captures essential aspects of quantum-gravitational effects on singularity formation and contributes to a deeper understanding of singularity resolution driven by quantum gravity.

Despite these limitations, the present work provides a concrete and internally consistent demonstration of how asymptotically safe quantum gravity effects can be incorporated into the WDW equation, and shows that, when present, classical black hole singularities can be quantum mechanically suppressed. Our results suggest that renormalization-group effects may play a fundamental role in shaping the quantum structure of spacetime at short distances, offering a novel and complementary mechanism for singularity resolution beyond interference-based interpretations. We hope that this study stimulates further investigations into the interplay between asymptotic safety, canonical quantum gravity, and the quantum physics of black hole interiors.

\bibliography{references}

\end{document}